TRB PAPER # 18-05614

# A Cooperative Freeway Merge Assistance System using Connected Vehicles

Md Salman Ahmed
Researcher, Center for Transportation Research, The University of Tennessee &
Vehicular Network Laboratory, East Tennessee State University
ahmedm@etsu.edu

Mohammad A Hoque, PhD
Director, Vehicular Network Laboratory, East Tennessee State University &
Researcher, Center for Transportation Research, The University of Tennessee
hoquem@etsu.edu

Jackeline Rios-Torres, PhD
Eugene P. Wigner Fellow, Oak Ridge National Laboratory
riostorresj@ornl.gov

Asad J. Khattak, Ph.D.
Beaman Professor, Department of Civil & Environmental Engineering
The University of Tennessee
akhattak@utk.edu

Word Count: 1241

ACKNOWLEDGEMENT

Support was provided by the US National Science Foundation under grant No. 1538139. Additional support was provided by the US Department of Transportation through the Collaborative Sciences Center for Road Safety, a consortium led by The University of North Carolina at Chapel Hill in partnership with The University of Tennessee and by the Laboratory Directed Research and Development Program of the Oak Ridge National Laboratory, Oak Ridge, TN 37831 USA, managed by UT-Battelle, LLC, for the DOE. Any opinions, findings, and conclusions expressed in this paper are those of the authors and do not necessarily reflect the views of the sponsors.

Submitted for Presentation Only to the
97th Annual Meeting Transportation Research Board
January 2018
Washington, D.C.

# A Cooperative Freeway Merge Assistance System using Connected Vehicles

**INTRODUCTION**

The rapid growth of traffic related fatalities and injuries around the world including developed countries has drawn researchers' attention for conducting research on automated highway systems to improve road safety over the past few years. In addition, fuel expenses due to traffic congestion in the U.S. translate to billions of dollars annually (*1*). These issues are motivating researchers across many disciplines to develop strategies to implement automation in transportation. The advent of connected-vehicle (CV) technology has added a new dimension to the research. The CV technology allows a vehicle to communicate with road-side infrastructure (vehicle-to-infrastructure), and other vehicles (vehicle-to-vehicle) on roads wirelessly using dedicated short-range communication (DSRC) protocol. Collectively, the vehicle-to-vehicle (V2V) and vehicle-to-infrastructure (V2I) communication technologies are known as V2X technology. Automotive companies have started to include On-Board Units (OBUs) on latest automobiles which can run safety-critical and assistive applications using V2X technology. For example, US Department of Transportation has already launched various applications including but not limited to lane-change assistance, collision avoidance, SPaT for emergency and transit vehicles. Merge conflicts, especially when vehicles are trying to merge from ramps to freeways, are a significant source of collisions, traffic congestion and fuel use (*2*, *3*). This paper describes a novel freeway merge assistance system utilizing V2X technology with the help of the DSRC protocol. The freeway merge assistance system uses an innovative three-way handshaking protocol and provides advisories to drivers to guide the merging sequence.

**METHODOLOGY**

Vehicles equipped with V2X capable OBUs transmit basic safety messages (BMSs) every tenth of a second. A BSM packet includes a vehicle's identifier, GPS positions, speed, time, and direction. The freeway merge assistance system utilizes these BSM packets and determines the optimal merging order for drivers using a 3-way handshaking communication protocol. Overall, the freeway merge assistance system performs the following steps to determine the merging order and generate advisories for drivers.

**Transmission and reception of BSM packets**. The transmission module of the freeway merge assistance system transmits the BSM, following the updating time of the GPS. The receiver module receives BSM packets from the surrounding vehicles within a range of 300~500 meters *(5)*. Both modules use a single-hop communication protocol (*6*) for transmitting and receiving the BSM packets.

**Observation of vehicular trajectories**. The freeway merge assistance system considers the ramp vehicles as the master vehicles who initiate all communication. A ramp vehicle triggers the core algorithm of the freeway merge assistance system as soon as the ramp vehicle enters into a ramp, notifying all the vehicles around notifying that a new vehicle entered the ramp. The data collection unit of the system collects and stores the vehicular trajectories of all the surrounding vehicles.

**Calculation of the merging order**. The data processing unit of the system filters out the noisy data from the collected trajectories and estimates the future trajectories to determine the merging point where all the ramp and freeway vehicles could meet. The unit also determines the approximated time for each vehicle to reach the merging point. Once the unit determines the time for each vehicle, the freeway merge assistance system initiates a three-way handshaking communication protocol (as illustrated in Figure 1) to transmit the timing information to all in range vehicles.

**Generation of advisory messages.** The freeway merge assistance system will then request a synchronization with the remaining vehicles to ensure they all in-range vehicles have the same timing information. Finally, when all the vehicles have mutually acknowledged and accepted the merging sequence, the freeway merge assistance system triggers the advisory generation module on each vehicle.

**Visualization of advisory message.** Once the merge assistance system is ready to generate the advisory messages on each vehicle, it will send the information to a map application on an Android device using Bluetooth connectivity. The messages will advise each driver according to the previously defined merging sequence and they will be displayed over a map marker with a text message. The application will also highlight the map marker (using a larger-size marker) for any referenced vehicle, e.g., if a vehicle on the freeway need to slow down to allow a ramp vehicle to merge first, the application will display the "Slow down" message to the freeway vehicle showing a larger-size map marker for the ramp vehicle on an Android device. Similarly, the application on an on-ramp vehicle may display the "merge behind" message showing a larger-size map marker for the freeway vehicle that the ramp vehicle will need to follow.

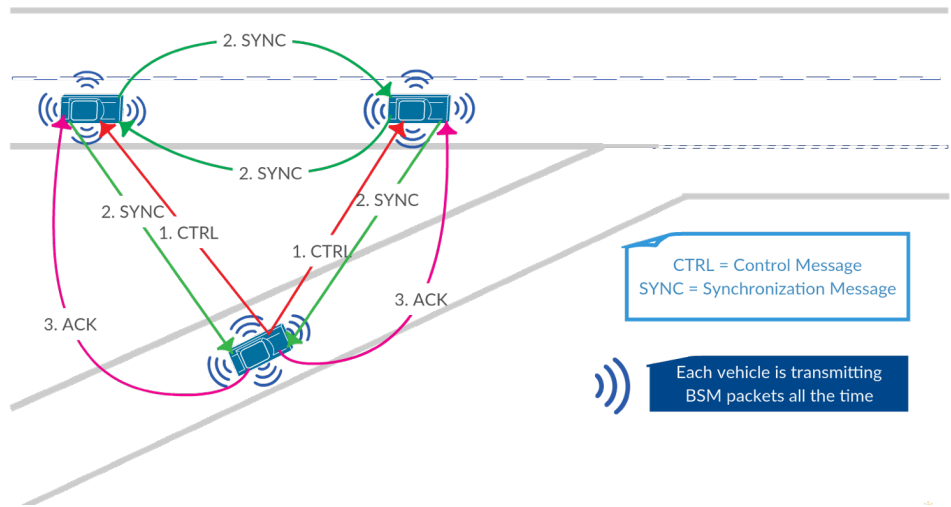

FIGURE 1 Three-way handshaking communication protocol

**EXPERIMENTAL RESULTS**

To evaluate the feasibility for real-time implementation of the freeway merge assistance system, we conducted field experiments on a real freeway. We chose exits 27, 32, 34, and 36— both East bound (EB) and West bound (WB) entrance ramps—on interstate I-26 near Johnson city, Tennessee. Three drivers who had valid US driver licenses and are frequent drivers of interstate highways participated in the experiment. Since the experiment involved human subjects (drivers), an official approval was obtained from the Institutional Review Board (IRB) of East Tennessee State University. The drivers received a training session to learn how the system works and how to interpret the advisory messages. To test the effectiveness of the system, two drivers were driving on the freeway while the third one entered the freeway using the ramp. To further test the capabilities of the system to handle conflicts, the on-ramp driver synchronized his entrance to the ramp with the lead vehicle on the freeway to merge at relatively the same time to generate a possible merge conflict. The second driver on the freeway followed the first driver while trying to keep a distance of around 50-100 meters. The merging process in this experiment was concluded successfully and the freeway merge assistance system was able to advise the drivers accordingly. Based on the results improvements are needed to ensure that the system provides advisory to all vehicles involved in the test and that ramp curvatures are accounted for in the calculation of distances and speeds.

## CONCLUSIONS

This paper details the development and real-world testing of a novel decentralized freeway merge assistance system. We evaluated the merge assistance system for eight exits along Interstate-26. The conducted experiments demonstrate that the system can successfully provide accurate advisory information for "diamond interchanges". The cooperative merging system can also generate large-scale data that can be used to fine-tune the system and study drivers' merging behavior in response to advisories. Future work will consider higher levels of vehicle automation and cooperative behavior in other scenarios.

## REFERENCES

1. D. Schrank, B. Eisele, T. Lomax, J. Bak, "2015 Urban Mobility Scorecard, Texas A&M Transportation Institute 2015" (2015).
2. S. Kato, Cooperative driving of autonomous vehicles based on localization, inter-vehicle communications and vision systems. *JSAE Rev*. **22**, 503–509 (2001).
3. J. Rios-Torres, A. . Malikopoulos, Automated and Cooperative Vehicle Merging at Highway On-Ramps. *IEEE Trans. Intell. Transp. Syst*. **18**, 780–789 (2017).
4. M. S. Ahmed, M. A Hoque, J. Rios-Torres, A. J. Khattak. Demo: Freeway Merge Assistance System using DSRC. ACM International Workshop on Connected and Automated Vehicle Mobility, ACM CarSys 2017.
5. D. Jordan, N. Kyte, S. Murray, M. A. Hoque, M. S. Ahmed, A. J. Khattak. Poster: Investigating Doppler Effects on Vehicle-to-Vehicle Communication: An Experimental Study. ACM International Workshop on Connected and Automated Vehicle Mobility, ACM CarSys 2017.
6. M. S. Ahmed, M. A. Hoque, A. J. Khattak, Demo: Real-time Vehicle Movement Tracking on Android Devices Through Bluetooth Communication with DSRC Devices. In *2016 IEEE Vehicular Networking Conference (VNC)* (2016), pp. 1–2.